\documentclass[12pt]{spieman} 
\usepackage{amsmath,amsfonts,amssymb}
\usepackage{pifont}
\usepackage{graphicx}
\usepackage{setspace}
\usepackage{tocloft}
\usepackage{textcomp}
\usepackage{color,soul}
\usepackage{subcaption}
\usepackage{float}
\usepackage[colorlinks=true, allcolors=blue]{hyperref}
\usepackage{multirow}
\usepackage[table,xcdraw,dvipsnames]{xcolor}
\usepackage{lineno}
\usepackage{makecell}
\usepackage{textcomp}
\usepackage{array}
\usepackage{listings}
\usepackage{mathptmx}
\usepackage{epsfig}
\usepackage{times}
\usepackage{float}
\usepackage{rotating}
\usepackage{makeidx}
\usepackage{url}
\usepackage{multirow}
\usepackage{booktabs}
\usepackage{tabularx}
\usepackage{blindtext}
\usepackage{adjustbox}
\usepackage{listings}
\usepackage{color}

\definecolor{dkgreen}{rgb}{0,0.6,0}
\definecolor{gray}{rgb}{0.5,0.5,0.5}
\definecolor{mauve}{rgb}{0.58,0,0.82}

\colorlet{Mycolor1}{green!10!orange}

\title{Glo-In-One-v2: Holistic Identification of Glomerular Cells, Tissues, and Lesions in Human and Mouse Histopathology}

\author[a]{Lining Yu}
\author[b]{Mengmeng Yin}
\author[a]{Ruining Deng}
\author[a]{Quan Liu}
\author[a]{Tianyuan Yao}
\author[a]{Can Cui}
\author[a]{Junlin Guo}
\author[b]{Yu Wang}
\author[c]{Yaohong Wang}
\author[d]{Shilin Zhao}
\author[b]{Haichun Yang}
\author[a,*]{Yuankai Huo}
\affil[a]{Vanderbilt University, Department of Computer Science, 2201 West End Ave, Nashville, USA, 37235}
\affil[b]{Vanderbilt University Medical Center, Department of Pathology, Microbiology and Immunology, 1211 Medical Center Dr, Nashville, USA, 37232}
\affil[c]{UT MD Anderson Cancer Center, Department of Anatomical Pathology, 1515 Holcombe Blvd, Unit 85 Houston, USA, 77030}
\affil[d]{Vanderbilt University Medical Center, Department of Biostatistics, 2525 West End Avenue, Suite 1100, Nashville, USA, 37232}

\cftpagenumbersoff{figure}
\cftpagenumbersoff{table} 
\begin{document} 
\begin{sloppypar}
\maketitle
\begin{abstract}

Segmenting glomerular intraglomerular tissue and lesions traditionally depends on detailed morphological evaluations by expert nephropathologists, a labor-intensive process susceptible to interobserver variability. Our group previously developed the Glo-In-One toolkit for integrated detection and segmentation of glomeruli. In this study, we leverage the Glo-In-One toolkit to version 2 with fine-grained segmentation capabilities, curating 14 distinct labels for tissue regions, cells, and lesions across a dataset of 23,529 annotated glomeruli across human and mouse histopathology data. To our knowledge, this dataset is among the largest of its kind to date. In this study, we present a single dynamic head deep learning architecture designed to segment 14 classes within partially labeled images of human and mouse pathology data. Our model was trained using a training set derived from 368 annotated kidney whole-slide images (WSIs) to identify 5 key intraglomerular tissues covering bowman’s capsule, glomerular tuft, mesangium, mesangial cells, and podocytes. Additionally, the network segments 9 glomerular lesion classes including adhesion, capsular drop, global sclerosis, hyalinosis, mesangial lysis, microaneurysm, nodular sclerosis, mesangial expansion, and segmental sclerosis. The glomerulus segmentation model achieved a decent performance compared with baselines, and achieved a 76.5\% average Dice Similarity Coefficient (DSC). Additional, transfer learning from rodent to human for glomerular lesion segmentation model has enhanced the average segmentation accuracy across different types of lesions by more than 3\%, as measured by Dice scores. The Glo-In-One-v2 model and trained weight have been made publicly available at \url{https://github.com/hrlblab/Glo-In-One_v2}.

\end{abstract}

{\noindent \footnotesize\textbf{*Corresponding Author:} Yuankai Huo,  \linkable{yuankai.huo@vanderbilt.edu} }

\begin{spacing}{2}  

\section{Introduction}

Whole-slide imaging (WSI) offers high-resolution tissue imaging, advancing quantitative assessments in nephropathology, particularly for glomerular analysis~\cite{huo2021ai}. Within renal pathology—a field marked by its distinct complexity in pathology image analysis—glomeruli are recognized as crucial functional units for clinical evaluations~\cite{jiang2021deep}.  To automate glomerular detection and segmentation, our team previously developed the Glo-In-One toolkit~\cite{yao2022glo} using deep learning with convolutional neural networks (CNNs). Despite its effectiveness, Glo-In-One is limited to whole-tuft segmentation without capturing sub-glomerular details.

\begin{figure*}[t]
\begin{center}
 \includegraphics[width=1.0\linewidth]{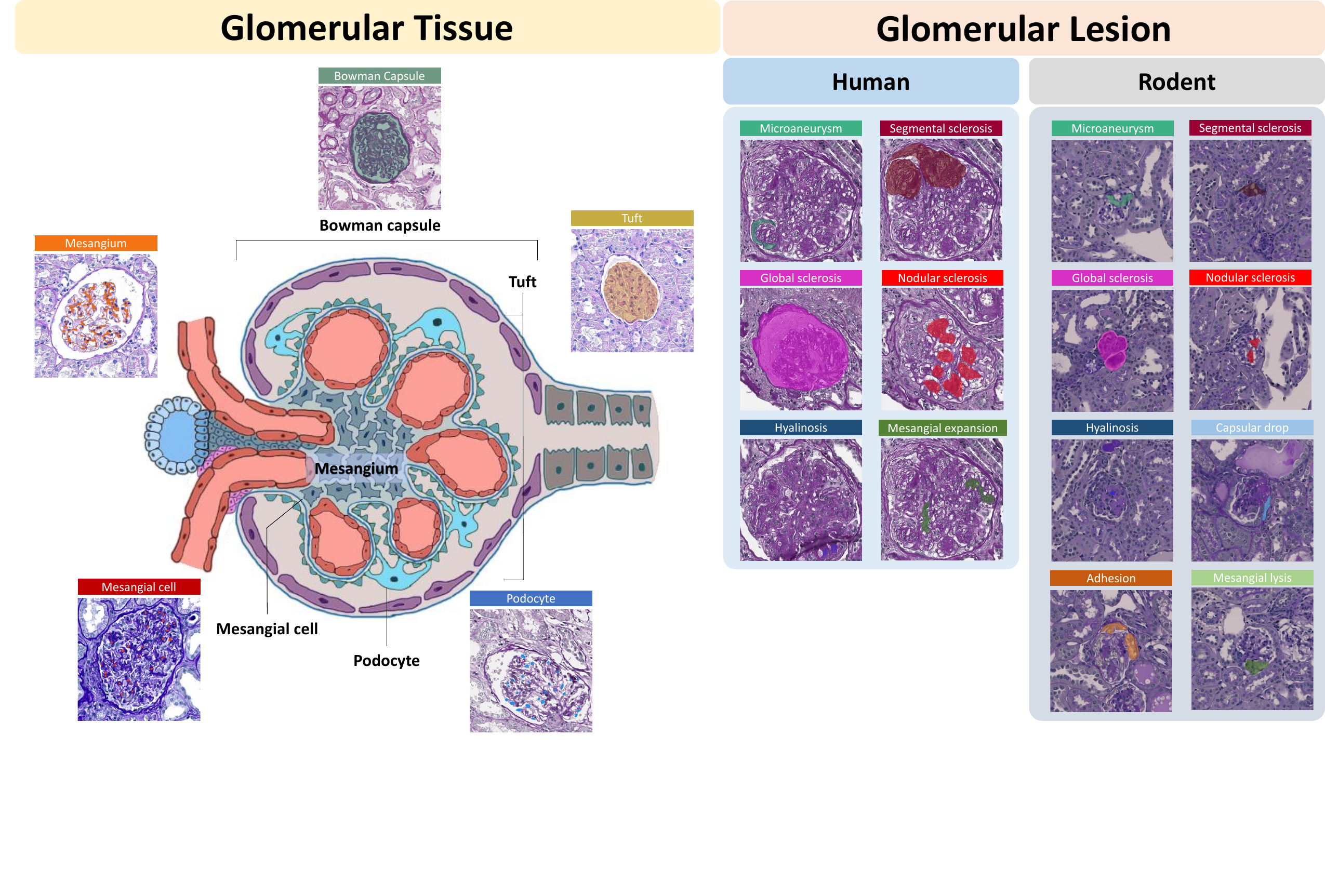}
 \vspace{-2.5cm}
\end{center}
   \caption{\textbf{Involved glomerular classes.} This figure presents fine-grained classes of intraglomerular tissue, including Bowman’s capsule (Cap), tuft (Tuft), mesangium (Mes), mesangial cells (Mec), and podocytes (Pod). It also highlights glomerular lesions observed in rodents and humans: adhesion (AH), capsular drop (CD), global sclerosis (GS), hyalinosis (HS), mesangial lysis (ML), microaneurysm (MA), nodular sclerosis (NS), mesangial expansion (ME), and segmental sclerosis (SS).} 
  \label{fig:types}
\end{figure*}

In this study, we introduce Glo-In-One-v2, an enhanced toolkit with fine-grained segmentation capabilities, applying 14 labels across tissue regions, cells, and lesions in a dataset of 23,529 annotated glomeruli — among the largest of its kind. Glo-In-One-v2 utilizes a dynamic deep learning architecture to segment 14 classes within partially labeled human and mouse pathology images (Fig.\ref{fig:types}). Trained on a training set split from 368 annotated kidney WSIs, the model identifies five internal renal structures (Bowman’s capsule, glomerular tuft, mesangium, mesangial cells, and podocytes) and segments nine glomerular lesion types, including adhesion, capsular drop, global sclerosis, hyalinosis, mesangial lysis, microaneurysm, nodular sclerosis, mesangial expansion, and segmental sclerosis.


Many recent studies have applied deep learning for glomerular quantification~\cite{janowczyk2016deep, komura2019machine,gadermayr2019cnn, esteva2019guide, wang2019pathology}. However, a limited focus has been placed on glomerular lesions within these studies. Glomerular lesions, markers of kidney damage, are crucial indicators of various kidney diseases~\cite{saikia2023mlp}. However, their small size and variability across observers make lesion segmentation more challenging than that of healthy glomeruli, and manual annotation is time-consuming, variable, and reliant on expert input, which limits reproducibility. Rodents, especially mice, are widely used in kidney disease research due to their genetic similarity to humans, affordability, small size, and rapid reproduction. Their short lifespan (1–2 years) allows researchers to study treatment effects and genetic modifications across generations, accelerating discoveries~\cite{kim2007comparative,smith2005modifications}. Furthermore, animal-to-human preclinical translation is essential for advancing drug development, therapies, diagnostics, and understanding disease mechanisms before clinical trials~\cite{brubaker2020translating,leenaars2019animal,ritskes2020improving,ritskes2020improving}. Thus, this study incorporates a glomerular lesion dataset from rodent kidneys to support cross-species glomerular segmentation. Our model is able to segment human samples by leveraging the shared similarities between rodent and human samples, allowing it to infer certain features present in human samples from rodent samples.

\begin{figure*}[t]
\begin{center}
 \includegraphics[width=1.0\linewidth]{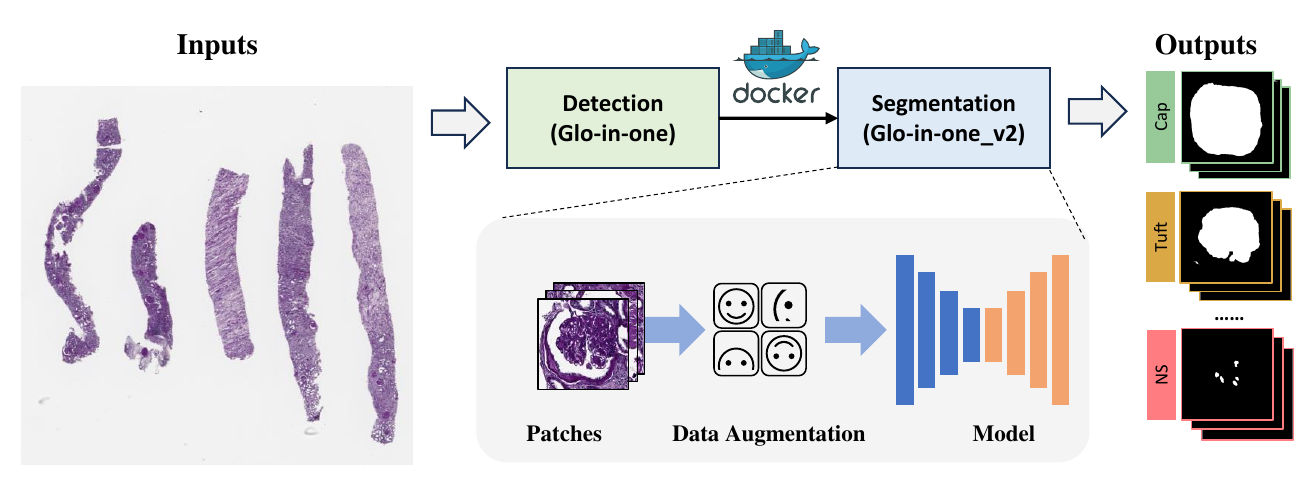}
\end{center}
   \caption{\textbf{Toolkit Overview.} This figure provides an overview of the Glo-In-One-v2 toolkit. The proposed toolkit is able to achieve 14 segmentation classes using a single command Docker command line. The input consists of raw WSIs, and the output is a holistic segmentation of glomeruli. The detection module, inherited from the previous toolkit version, delivers quantitative detection of glomeruli. The segmentation module utilizes a trained model, developed from patches extracted from WSIs with manual annotations provided by medical experts.}   
  \label{fig:toolkit}
\end{figure*}

\begin{figure*}[bth]
\begin{center}

    \begin{subfigure}{\textwidth}
        \centering
        \hspace{0.4cm}
        \includegraphics[width=0.8\textwidth]{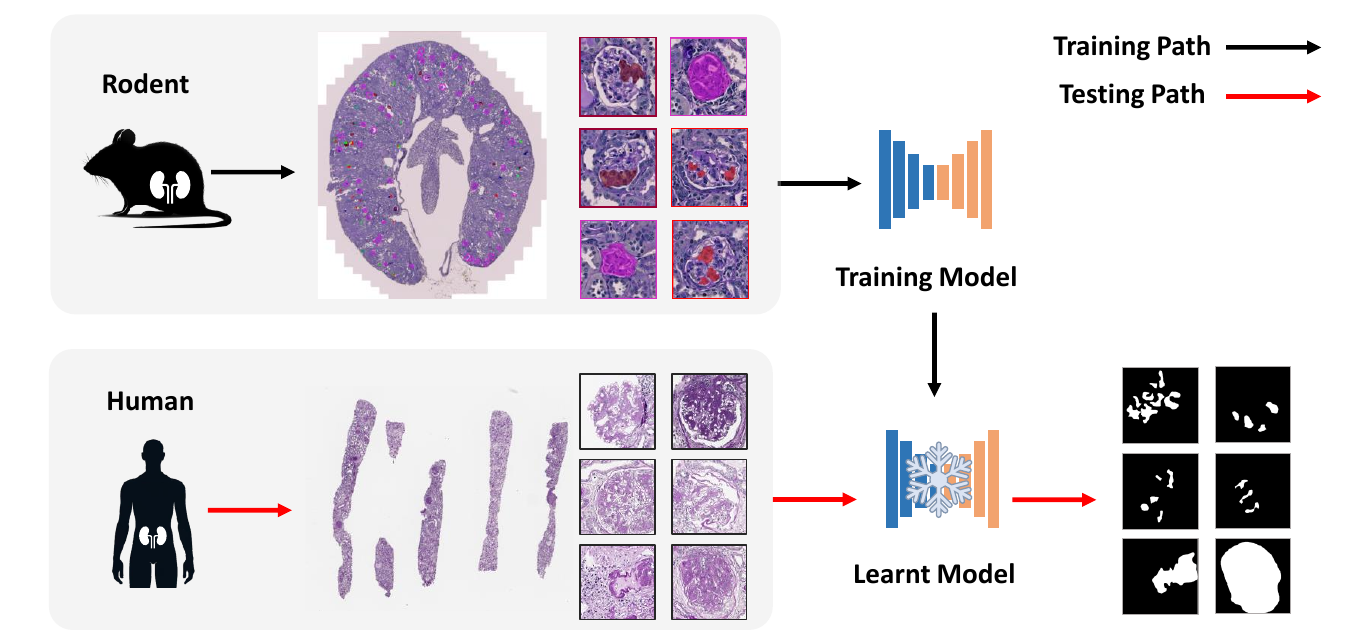}
        \caption{Rodent-to-Human Zero-shot Transfer Learning}
        \label{fig:zR2H}
    \end{subfigure}

    \begin{subfigure}{\textwidth}
        \centering
        \includegraphics[width=0.8\textwidth]{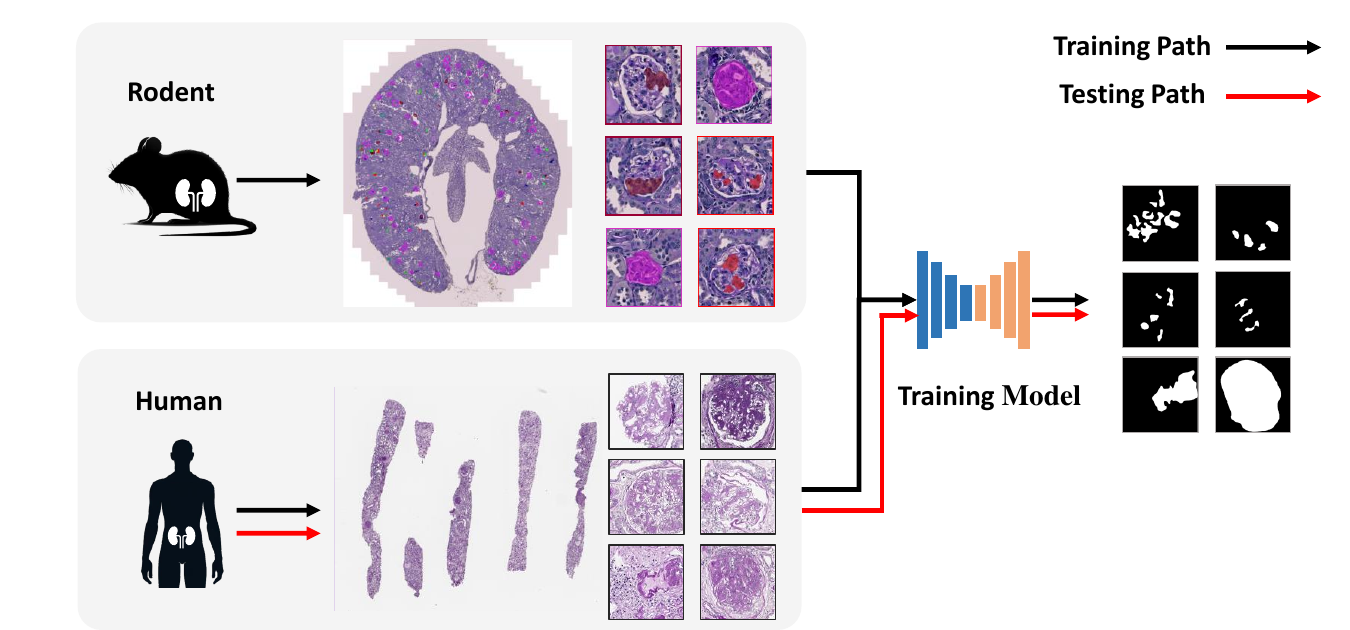}  
        \caption{Rodent-to-Human Hybrid Transfer Learning}
        \label{fig:RH2H}
    \end{subfigure}
\end{center}
   \caption{\textbf{Overview of the Rodent-to-human transfer learning framework.} This figure provides an overview of transfer learning for glomerular segmentation from rodent to human, where Fig \ref{fig:zR2H} illustrates the direct adaptation of a model trained on rodent data to human tasks without incorporating any knowledge from the human domain. In contrast. Fig \ref{fig:RH2H} demonstrates the use of a model that integrates knowledge learned from both rodent and human data for improved performance on human tasks. In the figures, the black arrows represent training paths, while the red arrows indicate testing paths.}   
  \label{fig:R2H}
\end{figure*}

Our contribution is fourfold: 
\begin{itemize}

    \item We introduce a single dynamic head deep learning architecture for comprehensive segmentation of 14 classes within partially labeled human and mouse pathology images across intraglomerular tissue and lesion.  
    \item We curated a dataset of over 23,000 glomeruli patches from 368 WSIs, offering detailed annotations for five intraglomerular tissue types (e.g., Bowman’s capsule, glomerular tufts) and nine glomerular lesions (e.g., global sclerosis, mesangial expansion, nodular sclerosis).
    \item We assess the feasibility of employing rodent data as an auxiliary source for human glomerular lesion segmentation, employing transfer learning to apply insights gained from rodent samples to human segmentation.
    \item We developed and released the open-source Glo-In-One-v2 toolkit (Fig.\ref{fig:types}), a containerized software solution for holistic glomerular segmentation. Built within a Docker container, it enables non-technical users to run the toolkit with a single command line. The toolkit is publicly accessible at \url{https://github.com/hrlblab/Glo-In-One_v2}.
    
\end{itemize}

\section{Related Work}
Several studies have explored the segmentation of various renal structures, including tubules, blood vessels, and interstitium, alongside the glomerulus ~\cite{hermsen2019deep,bouteldja2021deep}, as well as the components within the glomerulus itself ~\cite{ginley2019computational,zeng2020identification,kawazoe2022computational}, such as bowman’s capsule, glomerular tuft and mesangium. Accurately segmenting both the glomerulus and its intraglomerular tissue offers valuable insights into kidney diseases, facilitating the classification of pathological findings and enabling the development of prognostic models through the quantification of histopathological regions. However, the range of classes for intraglomerular tissue remains an area with potential for further expansion.
Besides, most current research on the glomerulus concentrates on patch-wise approaches, including the detection~\cite{bukowy2018region,yang2020circlenet} and segmentation~\cite{ginley2019computational} of numerous glomeruli within large patches, or binary classification within smaller patches~\cite{sheehan2019detection}. While there are studies aimed at classifying or identifying glomerular lesions, these efforts of nine require models to discern more complex features to accurately classify and differentiate between various types of lesions.

Previous works, such as the one by Nan et al.~\cite{nan2022automatic}, have introduced innovative methods for detailed recognition of glomerular lesions in kidney pathology, focusing on both segmentation and classification from WSIs. The Analytic Renal Pathology System (ARPS)~\cite{zeng2020identification} and the study by Akatsuka et al.~\cite{akatsuka2022automated} have made strides in automating the identification of glomerular lesions and cell types within kidneys using deep learning. However, these studies tend to cover a limited range of lesion types, which may overlook important pathological features, leading to a partial understanding of the lesions' complexity and variability. Moreover, segmentation tasks pose greater challenges than classification due to the need for a comprehensive grasp of context and detailed structures within the images. 

Our research expands on this foundation by incorporating a dataset comprising five intraglomerular tissue and nine glomerular lesions, with samples from both human and rodent WSIs. Additionally, we introduce a multi-label network optimized for training on this dataset to enhance prediction performance. Moreover, unlike traditional zero-shot strategies for rodent-to-human transfer learning, our approach leverages rodent data as support, as shown in Fig.\ref{fig:R2H}, enabling the network to further demonstrate its capability to predict glomerular lesions in human samples by utilizing knowledge gained during training on rodent data.

\section{Method}
\subsection{Segmentation Network}
\begin{figure*}[t]
\begin{center}
\includegraphics[width=1.0\linewidth]{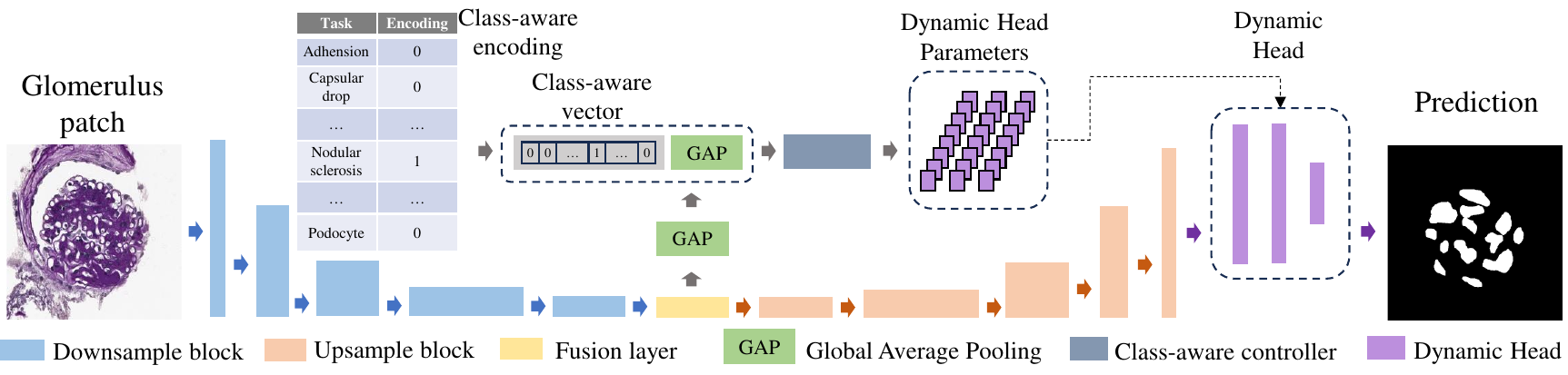}
\end{center}
\caption{{\textbf{Our pipeline.} This figure introduces the network structure of the dynamics head. Specifically, it contains a residual U-Net backbone, a class-aware controller, and a dynamic segmentation head.}}
\label{Framework}
\end{figure*}

We present our network, inspired from the research by~\cite{deng2022single},  a unified segmentation network that leverages a residual U-Net architecture to segment various glomerular classes within partially labeled images for pathology analysis.tailored for our intraglomerular tissue and lesion segmentation tasks.The network is detailed in Figure referred to as Fig.\ref{Framework}.

Unlike from fully labeled dataset, each image included in partially labeled dataset contains the annotations of only a specific class involved. We assign each segmentation on each class of tissue or lesion as a task for task-awareness for task-awareness, by encoding the task as a $m$-dimensional one-hot vector \cite{chen2017fast}. The $m$ is the number of classes, The encoding calculation for $T_k$, a class-aware vector of $i$th class of lesion is shown as follows:

\begin{equation}
T_k = \left\{
\begin{array}{ll}
1, & \text{if } k = i \\
0, & \text{\textit{otherwise}}
\end{array}
\right. \quad k = 1,2,...,m
\end{equation}

Dynamic filter generation \cite{zhang2021dodnet} was introduced to generate the kernels specialized to a particular class of segment tasks. The image feature $F$ is aggregated by a global average pooling (GAP) with $T_k$. The kernel parameters $\omega$ are computed as follows:

\begin{equation}
\omega = \phi(\text{GAP}(F)||T_k; \Theta_\phi)
\end{equation}

\noindent where $\Theta_\phi$ represents the controller parameters, ``$||$" represents the concatenation operation to combine high-level image features and the class-aware vector.$\phi$ reprensents a task-specific controller with a single 2D convolutional layer. 

The dynamic head was designed to achieve multi-label segmentation with three layers, denoted by $\omega_1$, $\omega_2$, $\omega_3$. The predictions of lesions can be generated as follows:

\begin{equation}
P = ((((M * \omega_1) * \omega_2) * \omega_3))
\end{equation}

\noindent where $ * $ is convolution, $M$ is the output from the decoder.

\subsection{Glomerular Mining with Glo-In-One}
We aim to obtain a large dataset of unannotated glomerular images from online sources such as open-access journals, the NIH Open-i$^\circledR$\cite{demner2012design} database, and general search engines to support self-supervised contrastive learning. Specifically, we collected 10,000 compound figures using keywords like ``glomerular OR glomeruli OR glomerulus" through the NIH Open-i$^\circledR$ search engine, with further details of the web image mining process available in~\cite{yao2021compound}.

However, images sourced online are typically embedded within compound figures containing multiple subplots, rendering them unsuitable for direct application in self-supervised learning. To address this, we utilized our previously developed compound image separation method~\cite{yao2021compound} to detect, isolate, and curate individual subplots as standalone images for subsequent learning tasks. Through this compound figure separation process, we obtained over 30,000 unannotated glomerular images from extensive web-based image mining.

\subsection{Containerization}

To facilitate glomerular quantification for non-technical users, we developed the Glo-In-Onev2 toolkit, show as in the Fig.~\ref{fig:toolkit}, an all-in-one solution that enables comprehensive glomerular detection and segmentation through a single, user-friendly command. By containerizing both the detection and segmentation modules within a Docker environment, we streamline the process so that users only need to input WSIs to obtain sophisticated multi-channel segmentation masks as output. Each channel in the mask is mapped to a specific intraglomerular tissue type or glomerular lesion class, offering a granular view tailored for in-depth analysis. This approach not only eliminates the need for specialized technical skills but also significantly reduces the time and effort required for advanced glomerular analysis, making it accessible and practical for a wider range of users, from clinical practitioners to researchers.

\section{Experiments and Results}

\subsection{Data Collection}

In this study, we curated a dataset comprising over 23,529 annotated glomerular patches obtained from 368 WSIs of renal pathology. Of these, as detailed in Table.\ref{table:data}, 16,943 patches were manually annotated by renal pathologists, while 6,586 patches were derived from the KPMP spatial segmentation dataset\cite{kpmp_data}. The annotations encompass intraglomerular tissue (Bowman’s capsule, tuft, and mesangium), cellular components (podocytes and mesangial cells), and a variety of glomerular lesions, including adhesion, capsular drop, global sclerosis, hyalinosis, mesangial expansion, mesangial lysis, microaneurysm, nodular sclerosis, and segmental sclerosis. All patches were extracted from WSIs scanned at the highest magnification originally and subsequently cropped and resized to a resolution of 512×512 pixels.

The dataset was stratified into training, validation, and test sets in a 6:1:3 ratio across all classes, ensuring patient-level splits to prevent data leakage, split result shown as detailed in Table.\ref{table:split}.


\begin{table}[h]
\centering{}
\caption{Distribution of subclasses in glomeruli}
\begin{adjustbox}{width=1.0\textwidth}
\begin{tabular}{c|ccccccccccccccccc}
\toprule
\multirow{2}{0.8in}{}  & \multicolumn{3}{c}{Region} & \multicolumn{2}{c}{Cell} & \multicolumn{9}{c}{Lesion} &  \multirow{2}{0.4in}{Total}\\
\cmidrule(lr){2-4}
\cmidrule(lr){5-6}
\cmidrule(lr){7-15}
&  Cap & Tuft & Mes & Prod & Mec & AH & CD & GS & HS & ME & ML & MA & NS & SS\\
\midrule
Rodent &  1,393 & 5,542 & 5,542 &  1,157 & 789 & 85 & 62 & 380 & 196 & -  & 71 & 203 & 342 & 196\\
Human & 6,586 & - & - & - & - & - & -  & 369 & 35 & 227 & - & 56 & 229 & 69\\
\midrule
Quantity & 7,979 & 5,542 & 5,542 &  1,157 & 789 & 85 & 62 & 749 & 231 & 227  & 71 & 259 & 571 & 265  & 23,529\\
\bottomrule
\end{tabular}
\label{table:data}
\end{adjustbox}
\end{table}

\begin{table}[!b]
\centering{}
\caption{Training set, Validation set and Testing set}
\begin{tabular}{c|ccc}
\toprule
  & Train & Val & Test \\
\midrule
Intraglomerular tissue  & 12,410 & 2,393   & 6,206  \\
Human lesion    & 513  & 200   & 272    \\
Rodent lesion & 969   & 159   & 407  \\
\bottomrule
\end{tabular}
\label{table:split}
\end{table}

\subsection{Experimental Design}

Utilizing an image pool concept inspired by Cycle-GAN~\cite{zhu2017unpaired}, we fed images into the model in batches of four, with the image pool size set matching the number of involved classes. When the number of images in the pool surpasses the batch size, a batch of images is selected for model input.

In the holistic segmentation task, the model was trained on the complete training set, encompassing all glomerular tissues and lesion types. For the rodent-to-human transfer learning task, we utilized our trained model, leveraging knowledge learned from both rodent and human data, to make predictions on both mouse and human test sets independently. For comparison, we defined three approaches as follows: (1) Human-to-human (H2H): a model trained on the human lesion training set and evaluated on the human test set; (2) Rodent-to-human (R2H): a model trained on the rodent lesion training set and evaluated on the human test set; (3) Rodent-and-human to human (R\&H2H): a model trained on a combined training set of rodent and human data, limited to lesion data only, and evaluated on the human test set.

To evaluate the model's performance, we employed metrics such as the Dice similarity coefficient (Dice). The best-performing model on the validation dataset was chosen based on the average Dice score over 200 epochs, and its performance was then evaluated using this model. Our experiments were conducted using an NVIDIA RTX A5000 with 24G of VRAM.

We compared the introduced network to baseline models, including (1) multiple individual U-Net models (U-Nets) ~\cite{ronneberger2015u}, (2) multiple individual DeepLabv3 models (DeepLabv3s)~\cite{lutnick2019integrated}, and (3) a multi-class segmentation model for partially
labeled datasets~\cite{gonzalez2018multi} for renal pathology quantification. Additionally, the performance of the network was evaluated against transformer baselines (4) a hybrid neural network architecture that combines the Swin Transformer with the U-Net transformer encoder for enhanced medical image segmentation (Swin-Unetr)~\cite{hatamizadeh2021swin}, and (5) an approach to semantic segmentation based on the Vision Transformer  (Segmenter)~\cite{strudel2021segmenter}. All of the parameter settings are followed by original paper.

\subsection{Results}
\subsubsection{Holistic Segmentation}

\begin{figure*}[t!]
\begin{center}

 \includegraphics[width=0.8\linewidth]{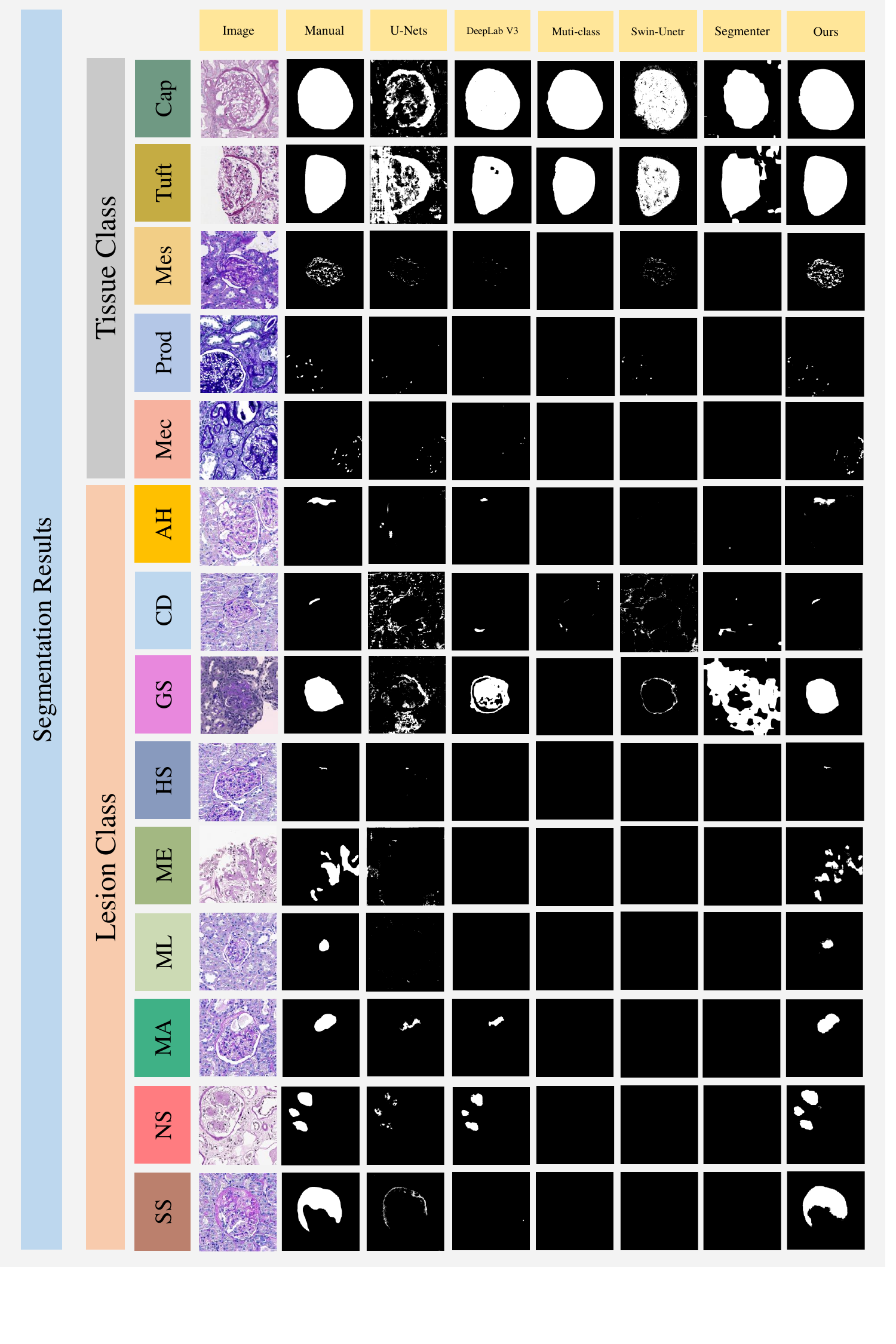}
 
\end{center}
   \vspace{-1.0cm}
   \caption{\textbf{Qualitative Results of different methods on glomerular segmentation.} This figure displays the qualitative outcomes of various segmentation methods for all classes of glomeruli. The first column features the original, unannotated images, while the second column shows the manual segmentation results.}
  \label{fig:Quali_all}
\end{figure*}

As shown in Table \ref{table:all_result},the performance metrics for the segmentation of each class of glomerular tissue and lesion over the entire rodent dataset. And Fig.\ref{fig:Quali_all} presents the qualitative result about the performance of different methods on the multi-label dataset. The experimental results show that our trained model as a single multi-label model can achieve better performance on prediction classes of both glomerular tissues and lesions than baseline methods. such as CNNs-based(i.e.DeepLabV3) and Transformers-based (i.e.Swin-unetr).

The results suggest that, although multi-head architectures face challenges in capturing spatial relationships between objects (e.g., subset-superset associations between the bowman capsule and tuft), the introduced dynamic-head approach outperforms other methods.

\subsubsection{Rodent-to-Human Transfer Learning}

As shown in Table \ref{table:2H_result}, we evaluated the predictive accuracy of models trained solely on rodent samples across various common lesion categories in human subjects. In the Human-to-human (H2H) experiments, the limited availability of human samples proved insufficient for the model to acquire the necessary knowledge to predict certain classes accurately. When we opted to apply rodent models directly to human prediction (R2H) in a zero-shot transfer learning manner, performance was suboptimal due to morphological differences and other domain gaps. However, using the hybrid transfer learning strategy (R\&H2H), where rodent samples serve as an auxiliary source for human sample prediction, performance was enhanced when our model is able to acquire knowledge more comprehensively from both human and rodent data by leveraging the decent performance on the rodent data, as detailed in Table \ref{table:M2M_result} in the appendix. 

Moreover, despite potential disturbances from tissue classes with superset relationships (e.g., Bowman’s capsule tissue and glomerular lesions), our model demonstrated robust performance, effectively overcoming these challenges.

\begin{figure*}[t!]
\begin{center}
 \includegraphics[width=0.8\linewidth]{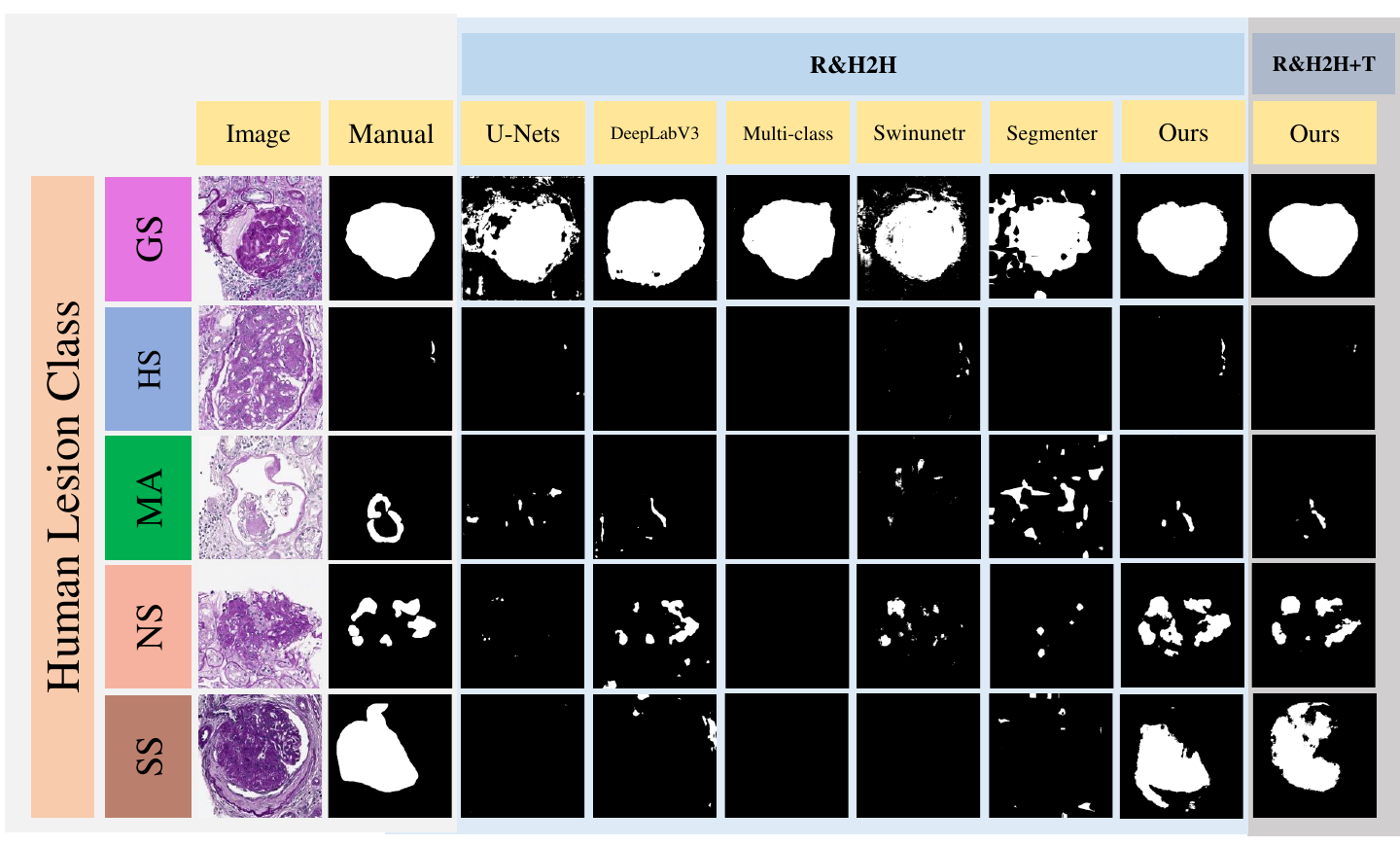}
\end{center}
   \caption{\textbf{Qualitative Results of different methods using the hybird tranfer learning stragety on human glomerular lesions.} This figure displays the qualitative outcomes of various segmentation methods for all common classes of glomerular lesions. The first column features the original, unannotated images, while the second column shows the manual segmentation results.Subsequent columns belong to two section: "R\&H2H" and "R\&H2H+T".}
  \label{fig:Quali_hum}
\end{figure*}

Additionally, to evaluate the performance of models on human samples and compare the effect of transfer learning, Table \ref{table:2H_result} presents performance metrics for all lesion classes in the complete human dataset. Figure \ref{fig:Quali_hum} provides a qualitative comparison of the performance of different methods in adapting rodent models for prediction tasks on the human dataset.

\begin{table*}[bth]
\caption{Performance of different models for glomerular tissue and lesion segmentation. Dice similarity coefficient (\%, the higher, the better) is used for evaluation. The bold mark indicates the best performance.}
\begin{center}
\begin{adjustbox}{width=1.0\textwidth}
\begin{tabular}{l|c|ccccccccc}
\toprule
\multirow{2}{0.8in}{Method} & \multirow{2}{0.8in}{Backbone} & \multicolumn{3}{c}{Region} & \multicolumn{2}{c}{Cell} & \multicolumn{3}{c}{Lesion} \\
\cmidrule(lr){3-5}
\cmidrule(lr){6-7}
\cmidrule(lr){8-10}
& &  Cap & Tuft & Mes & Prod & Mec & AH & CD & GS \\
\midrule
U-Nets~\cite{ronneberger2015u} & CNN & 74.6 & 59.4 & 59.8 & 71.1  & 60.9  & 50.2 & 50.8 & 46.0 \\
DeepLabV3~\cite{lutnick2019integrated} & CNN & 82.9 & 67.3 & 51.8 & 58.4 & 50.9 & 50.2 & 58.7 & 71.1 \\
Multi-class~\cite{gonzalez2018multi} & CNN & 95.0 & 46.9 & 49.3  & 49.9 & 49.9 & 49.7 & 59.0 & 44.3 \\
\midrule
Swinunetr~\cite{hatamizadeh2021swin}  & Transformer & 82.9 & 71.1 & 73.4  & 69.8 & 50.1 & 50.2 & 52.3 & 47.7 \\
Segmenter~\cite{strudel2021segmenter}  & Transformer & 72.8 & 72.0 & 49.4 & 50.5 & 50.9 & 46.7 & 55.8 & 55.7 \\
\midrule
Ours & CNN & \textbf{96.3} &  \textbf{97.0}  & \textbf{89.5} & \textbf{76.4} & \textbf{66.6} & \textbf{59.6} & \textbf{67.5} & \textbf{93.6} \\
\bottomrule

\end{tabular}
\end{adjustbox}

\begin{adjustbox}{width=1.0\textwidth}
\begin{tabular}{l|c|ccccccc}
\toprule
\multirow{2}{0.8in}{Method} & \multirow{2}{0.8in}{Backbone} & \multicolumn{6}{c}{Lesion} &  \multirow{2}{0.4in}{Average}\\
\cmidrule(lr){3-8}
& & HS & ME & ML & MA & NS & SS & \\
\midrule

U-Nets~\cite{ronneberger2015u} & CNN & 63.4 & 44.1 & 49.5 & 60.2 & 52.4 & 52.7 & 56.8\\
DeepLabV3~\cite{lutnick2019integrated} & CNN & 56.9 & 49.1 & 50.4 & 56.2 & 64.0 & 50.2 & 58.4 \\

Multi-class~\cite{gonzalez2018multi} & CNN &  49.9 & 49.1 & 49.6 & 49.3 & 49.0 & 48.1 & 52.8 \\
\midrule
Swinunetr~\cite{hatamizadeh2021swin}  & Transformer & 50.3 & 49.1 & 49.6 & 51.3 & 49.5 & 48.2 & 56.8\\
Segmenter~\cite{strudel2021segmenter}  & Transformer  & 50.5 & 50.0 & 49.6 & 55.5 & 51.2 & 50.2& 54.3 \\
\midrule
Ours  & CNN & \textbf{73.2} & \textbf{66.2} &\textbf{ 57.1} & \textbf{71.5} & \textbf{76.6} & \textbf{79.4} & \textbf{76.5}\\
\bottomrule

\end{tabular}
\end{adjustbox}
\end{center}
\label{table:all_result}
\end{table*}

\begin{table*}[h!]
\caption{Performance of different model and strategy: Human-to-human (H2H), Rodent-to-Human (R2H), Rodent\&Human-to-human (R\&H2H) and Rodent\&Human-to-human plus additional tissue data (R\&H2H+T) . Dice similarity coefficient (\%, the higher, the better) is used for evaluation. The bold mark indicates the best performance.}

\begin{center}
\begin{tabular}{l|c|ccccccccc}
\toprule
\multirow{2}{0.8in}{Method} & \multirow{2}{0.8in}{Approach}& \multicolumn{5}{c}{Human Glomerular Lesion} & \multirow{2}{0.8in}{Average}\\
\cmidrule(lr){3-7}
&& GS & HS & MA & NS & SS \\
\midrule
U-Nets~\cite{ronneberger2015u} 
& H2H & 91.4 & 49.9 & 49.4 & 67.0 & 47.8 & 61.1\\
& R2H  & 70.6 & 49.9 & 50.9  & 67.8  & 47.8  & 57.4 \\
& R\&H2H  & 65.9  & 55.2 & 55.6 & 59.0 & 47.8 & 56.7\\
& R\&H2H+T & 39.7  & 51.0 & 52.9  & 53.2 & 52.9 & 49.9\\
DeepLabV3~\cite{lutnick2019integrated}  
& H2H   & 95.0  & 54.1 & 55.7 & 70.9 & 52.1 & 65.6\\
& R2H & 78.2 & 52.7  & 49.9  & 56.4  & 60.8  & 59.6 \\
& R\&H2H  & 74.8  & 54.1 & 54.1 & 68.5 & 50.7 & 60.4 \\
& R\&H2H+T  & 79.3 & 52.7   & 51.3 & 60.6 & 47.8 & 58.3\\
Multi-class~\cite{gonzalez2018multi} 
& H2H  & 92.6 & 49.9  & 49.4 & 48.4 & 47.8 & 57.6\\
& R2H & 78.0 & 49.9 & 49.4 & 48.4  & 47.8  & 54.7\\
& R\&H2H  & 91.4  & 49.9 & 49.4 & 48.4 & 47.8 & 57.4\\
& R\&H2H+T  & 41.0  & 50.0 & 49.4 & 48.3 & 47.8 & 47.3\\
Swinunetr~\cite{hatamizadeh2021swin} 
& H2H  & 90.4 & 49.9 & 49.5 & 65.0 & 47.8 & 60.5 \\
& R2H & 81.5 & 50.1 & 49.5 & 53.3  & 49.6  & 56.8 \\
& R\&H2H  & 79.3 & 53.4 & 49.4 & 63.3 & 47.8 & 58.6\\
& R\&H2H+T & 44.1 & 49.9 & 49.4 & 48.4 & 47.8 & 47.9\\
Segmenter~\cite{strudel2021segmenter} 
& H2H   & 83.3 & 50.2 & 50.6 & 49.2 & 48.0 & 56.3 \\
& R2H & 82.5 & 50.4 & 50.7 & 49.4 & 49.4 & 56.5 \\
& R\&H2H & 66.6  & 49.9 & 50.0 & 55.9 & 48.4 & 54.2\\
& R\&H2H+T & 61.6 & 49.9  & 49.4 & 50.4 & 47.2 & 51.7\\
\bottomrule
Ours 
& H2H  & 94.3 & 49.9 & 50.5 &  \textbf{76.5 }& 68.5 & 67.9\\
& R2H  & 79.3 & 49.9  & 50.5  & 67.8  & 69.5 & 63.4\\
& R\&H2H  
& 94.0  & \textbf{57.7 }& 55.3  & 75.3 & 69.6 & 70.4 \\
& R\&H2H+T
& \textbf{95.2} & 54.5  & \textbf{58.7 }& 76.1  & \textbf{71.3}  & \textbf{71.2} \\
\bottomrule

\end{tabular}

\end{center}
\label{table:2H_result}
\end{table*}

\section{Discussion}

As shown in the Figure \ref{fig:types}, it is apparent that some classes in our glomeruli dataset are not completely mutually exclusive but exhibit relationships such as overlap, subset, or superset. For example, in tissue classes, the Bowman’s capsule region of a glomerulus can contain the tuft region, and the tuft region can further contain the mesangium. Another case is the lesion class of global sclerosis, where the mask region covers almost the entire Bowman’s capsule with significant overlap. Traditional multi-head models struggle with these relationships because they tend to segment the image into separate channels, lacking strong associations between them. These associations have a notable impact on the performance of the involved classes.

According to the experimental results, as shown in Table \ref{table:2H_result}, multi-head baseline models, particularly for global sclerosis, show reduced performance when incorporating the tissue training set (R\&H2H+T) compared to when trained exclusively on lesion data (R\&H2H). This indicates that the addition of tissue data can undermine the model's effectiveness in these cases.However, our introduced model with dynamic heads maintains robust performance even when trained on a diverse set of tissue and lesion classes, including those with overlapping or hierarchical relationships, such as subset or superset structures. As presented in Table \ref{table:all_result}, our model achieves strong segmentation results across all glomerular tissue and lesion classes, outperforming all baseline methods.

In the rodent-to-human glomerular lesion transfer learning task, we evaluated the performance of models trained exclusively on rodent data, models trained solely on human data, and models trained on a combined dataset of both rodent and human data. Despite the inherent domain shift and the influence of glomerular tissue classes, potentially associated with lesion classes, present in the training data, our model consistently outperforms all baseline methods. These results highlight the effectiveness of leveraging knowledge from rodent datasets to enhance the model’s ability to accurately identify and segment glomerular lesions in human samples.

\section{Conclusions}

In this paper, we develop and release a holistic Glo-In-One-v2 open-source toolkit to provide holistic identification of glomerular glomerular cells, tissues and lesion in human and mouse histopathology. The containerized machine learning system process renal WSIs in a fully automated manner with a single command line, which is user-friendly for non-technical users. Our curated dataset is extensive, covering five tissue classes and nine lesion classes from both rodent and human pathology. Experimental results demonstrate that our model outperforms baseline methods in segmenting tissues and lesions, even in cases where classes overlap or exhibit subset or superset relationships. To assess transfer learning from rodent to human, we compared models trained solely on rodent data with those trained exclusively on human data. The findings show that our holistic model achieves robust performance across datasets, highlighting its generalizability.



\appendix    

\section{Rodent-to-Rodent Supervised Learning}

In transfer learning, ensuring that a model performs well on the source domain is essential, as this capacity forms the foundation for successful adaptation to a target domain. For our study, the model’s proficiency in identifying rodent lesions was evaluated rigorously to validate its predictive robustness for applications in human lesion analysis. Specifically, we trained our models exclusively on rodent domain data, which allowed us to carefully assess their effectiveness in distinguishing intricate lesion patterns. Experimental results indicate that our model consistently outperforms all baseline methods on the rodent dataset, suggesting a higher level of feature extraction and prediction capabilities, establishing a critical foundation for effective transfer learning from rodent to human lesion identification.

\begin{table*}[!h]
\caption{Performance of different model trained by rodent lesion data for prediction on rodent glomerular lesion segmentation. Dice similarity coefficient (\%, the higher, the better) is used for evaluation. The bold mark indicates the best performance.}

\begin{center}
\begin{tabular}{l|cccccccccc}
\toprule
\multirow{2}{0.8in}{Method} & \multicolumn{8}{c}{Rodent Glomerular Lesion} & \multirow{2}{0.8in}{Average}\\
\cmidrule(lr){2-9}

& AH & CD & GS & HS & ML & MA & NS & SS \\
\midrule
U-Nets~\cite{ronneberger2015u} 
& 49.7 & 64.7 & 81.5 & 70.5 & 49.4 & 71.6 & 62.3 & 47.4 & 62.1\\
DeepLabV3~\cite{lutnick2019integrated} 
& 54.5 & 66.6 & 87.9 & 68.5 & 52.0 & 69.5 & 70.1 & 66.0 & 66.9\\
Multi-class~\cite{gonzalez2018multi} 
& 49.7 & 49.6 & 87.6 & 49.9 & 49.4 & 49.4 & 49.2 & 47.4 & 54.0\\
Swinunetr~\cite{hatamizadeh2021swin} 
& 50.3 & 63.3 & 84.3 & 73.3- & 50.8 & 70.5 & 66.8 & 49.3 & 63.6\\
Segmenter~\cite{strudel2021segmenter} 
& 49.7 & 51.4 & 79.8 & 50.8 & 50.2 & 66.8 & 56.5 & 51.5 & 57.1\\
\bottomrule
Ours 
& \textbf{58.7} & \textbf{68.1} & \textbf{89.7} & \textbf{75.5} & \textbf{57.2} & \textbf{76.9} & \textbf{77.9 }& \textbf{77.7} & \textbf{72.7}\\
\bottomrule

\end{tabular}

\end{center}
\label{table:M2M_result}
\end{table*}






\section*{Disclosures}
The authors declare that there are no financial interests, commercial affiliations, or other potential conflicts of interest that could have influenced the objectivity of this research or the writing of this paper.

\section* {Code, Data, and Materials Availability} 
The code used in this study is publicly available at \url{https://github.com/hrlblab/Glo-In-One_v2}. However, a portion of the data used in this research includes datasets obtained from Vanderbilt University Medical Center (VUMC). Due to privacy and institutional policies, data sharing requires permission from VUMC and a data use agreement. Therefore, the data cannot be made publicly available at this time. Interested researchers may contact VUMC to explore potential data access options.

\section* {Acknowledgments}
This research was supported by NIH R01DK135597(Huo), DoD HT9425-23-1-0003(HCY), NIH NIDDK DK56942(ABF). This work was also supported by Vanderbilt Seed Success Grant, Vanderbilt Discovery Grant, and VISE Seed Grant. This project was supported by The Leona M. and Harry B. Helmsley Charitable Trust grant G-1903-03793 and G-2103-05128. This research was also supported by NIH grants R01EB033385, R01DK132338, REB017230, R01MH125931, and NSF 2040462. We extend gratitude to NVIDIA for their support by means of the NVIDIA hardware grant. This works was also supported by NSF NAIRR Pilot Award NAIRR240055.

The KPMP is funded by the following grants from the NIDDK: U01DK133081, U01DK133091, U01DK133092, U01DK133093, U01DK133095, U01DK133097, U01DK114866, U01DK114908, U01DK133090, U01DK133113, U01DK133766, U01DK133768, U01DK114907, U01DK114920, U01DK114923, U01DK114933, U24DK114886, UH3DK114926, UH3DK114861, UH3DK114915, UH3DK114937. The content is solely the responsibility of the authors and does not necessarily represent the official views of the National Institutes of Health.

Language and grammar clean-up of this manuscript was facilitated through the use of ChatGPT, an AI language model developed by OpenAI.
 

\bibliography{main}   
\bibliographystyle{spiejour}   


\vspace{2ex}\noindent\textbf{First Author} is currently a MS student in Computer Science and at Vanderbilt University. He is supervised by Prof. Yuankai Huo at HRLB Lab. His main research interests include medical image analysis, deep learning, and computer vision.

\vspace{1ex}
\noindent Biographies and photographs of the other authors are not available.

\listoffigures
\listoftables

\end{spacing}
\end{sloppypar}
\end{document}